\documentclass[aps,prb,twocolumn,showpacs,superscriptaddress,groupedaddress]{revtex4-2}

\usepackage{graphicx}
\usepackage{dcolumn}
\usepackage{bm}
\usepackage{xcolor}

\begin{document}

\title{Spatially inhomogeneous superconductivity in UTe$_{2}$}

\author{S. M. Thomas$^{1}$, C. Stevens$^{2}$, F. B. Santos$^{3}$,  S. S. Fender$^{1}$, E. D. Bauer$^{1}$, F. Ronning$^{1}$, J. D. Thompson$^{1}$, A. Huxley$^{2}$, and P. F. S. Rosa$^{1}$}
\affiliation{
$^{1}$  Los Alamos National Laboratory, Los Alamos, New Mexico 87545, U.S.A.\\
$^{2}$ School of Physics and Astronomy, University of Edinburgh, Edinburgh, UK.\\
$^{3}$ Escola de Engenharia de Lorena, Universidade de Sao Paulo (EEL-USP), Materials Engineering Department (Demar), Lorena, Sao Paulo, Brazil.}
\date{\today}

\begin{abstract}
Newly-discovered superconductor UTe\(_2\) is a strong contender for a topological spin-triplet state wherein a multi-component order parameter arises from two nearly-degenerate superconducting states.
A key issue is whether both of these states intrinsically exist at ambient pressure. 
Through thermal expansion and calorimetry, we show that UTe$_2$ at ambient conditions exhibits two detectable transitions only in some samples, and the size of the thermal expansion jump at each transition varies when the measurement is performed in different regions of the sample.
This result indicates that the two transitions arise from two spatially separated regions that are inhomogeneously mixed throughout the volume of the sample, each with a discrete superconducting transition temperature (T$_c$).
Notably, samples with higher T$_c$ only show a single transition at ambient pressure.
Above 0.3~GPa, however, two transitions are invariably observed in ac calorimetry.
Our results not only point to a nearly vertical line in the pressure-temperature phase diagram but also provide a unified scenario for the sample dependence of UTe$_{2}$.
\end{abstract}

\maketitle

UTe$_2$ is a recently discovered superconductor that exhibits many intriguing properties.
Even though UTe$_2$ does not exhibit long-range magnetic order above 25~mK, initial reports placed UTe$_2$ as a new example of a spin-triplet superconductor due to an upper critical field (H$_{c2}$) exceeding 30~T and scaling of the magnetization indicating proximity to a ferromagnetic quantum critical point~\cite{Ran2019, Aoki2019, Sundar2019}.
Importantly, superconductivity in UTe$_2$ may be topological.
Asymmetric tunneling was observed across step edges in scanning tunneling microscopy, consistent with chiral superconductivity~\cite{Jiao2019}.
Polar Kerr effect measurements combined with theoretical modelling revealed that the superconducting order parameter breaks time-reversal symmetry and is likely to contain Weyl nodes~\cite{Hayes2020}.
More recently, magnetic penetration depth measurements revealed temperature scaling consistent with a multi-component spin-triplet state~\cite{Ishihara2021}.

UTe$_2$ also exhibits striking phase diagrams as a function of applied pressure and magnetic fields.
For instance, re-entrant superconductivity is observed for field applied in the orthorhombic $bc$ plane, whereas a metamagnetic transition occurs near 30~T for fields parallel to the $b$~axis~\cite{Ran2019a,Knebel2019,Knafo2020}.
Under pressure, UTe$_2$ remains equally puzzling, and a complete agreement between the many reports has yet to be reached. One common aspect is the existence of multiple superconducting transitions under pressures above about 0.3~GPa~\cite{Braithwaite,Ran2019b,Knebel2020,Aoki2020,Thomas2020}.
One superconducting transition (T$_{c1}$) reaches a maximum of about 3~K at a pressure near 1.2~GPa, and a second superconducting transition (T$_{c2}$) is suppressed monotonically with pressure.
Above 1.2~GPa, T$_{c1}$ is rapidly suppressed and a new non-superconducting ordered phase emerges.
Though this phase was initially thought to be the ferromagnetic state responsible for fluctuations leading to spin-triplet superconductivity at zero pressure, more recent reports argue for antiferromagnetic order under pressure due to the presence of two magnetic phase transitions as a function of temperature and their suppression as a function of all applied field directions~\cite{Aoki2020,Thomas2020}.
Magnetic susceptibility and magnetization measurement under pressure provided further support for antiferromagnetic order above 1.2~GPa~\cite{Li2021}.
Neutron measurements found that inelastic scattering is dominated by incommensurate spin fluctuations~\cite{Duan2020,Knafo2021} in spite of muon spin resonance and nuclear magnetic resonance experiments arguing for ferromagnetic fluctuations~\cite{Sundar2019,Tokunaga2019}.
It was later argued that antiferromagnetic fluctuations may be responsible for superconductivity in UTe$_2$~\cite{Duan2021}. 

A key point of contention is the low-pressure region of the phase diagram.
Whether two superconducting transitions exist at ambient pressure or inhomogeneities drive a split transition remains an open question.
On one hand, two nearby transitions were observed in heat capacity by Hayes \textit{et al}, and the Kerr effect sets in only at temperatures below the lower temperature transition~\cite{Hayes2020}.
On the other hand, a composition dependence study argued that the highest quality samples only show a single transition in heat capacity~\cite{Cairns2020}.
A two-component order parameter, however, is necessary for the proposed Weyl superconductivity and non-zero Kerr effect.
Because of the orthorhombic structure of UTe$_2$, there is no underlying symmetry argument for the existence of a two-component order parameter, and the existence of two nearby transitions would therefore be accidental.

\begin{figure*}[!ht]
    \includegraphics[width=1.0\textwidth]{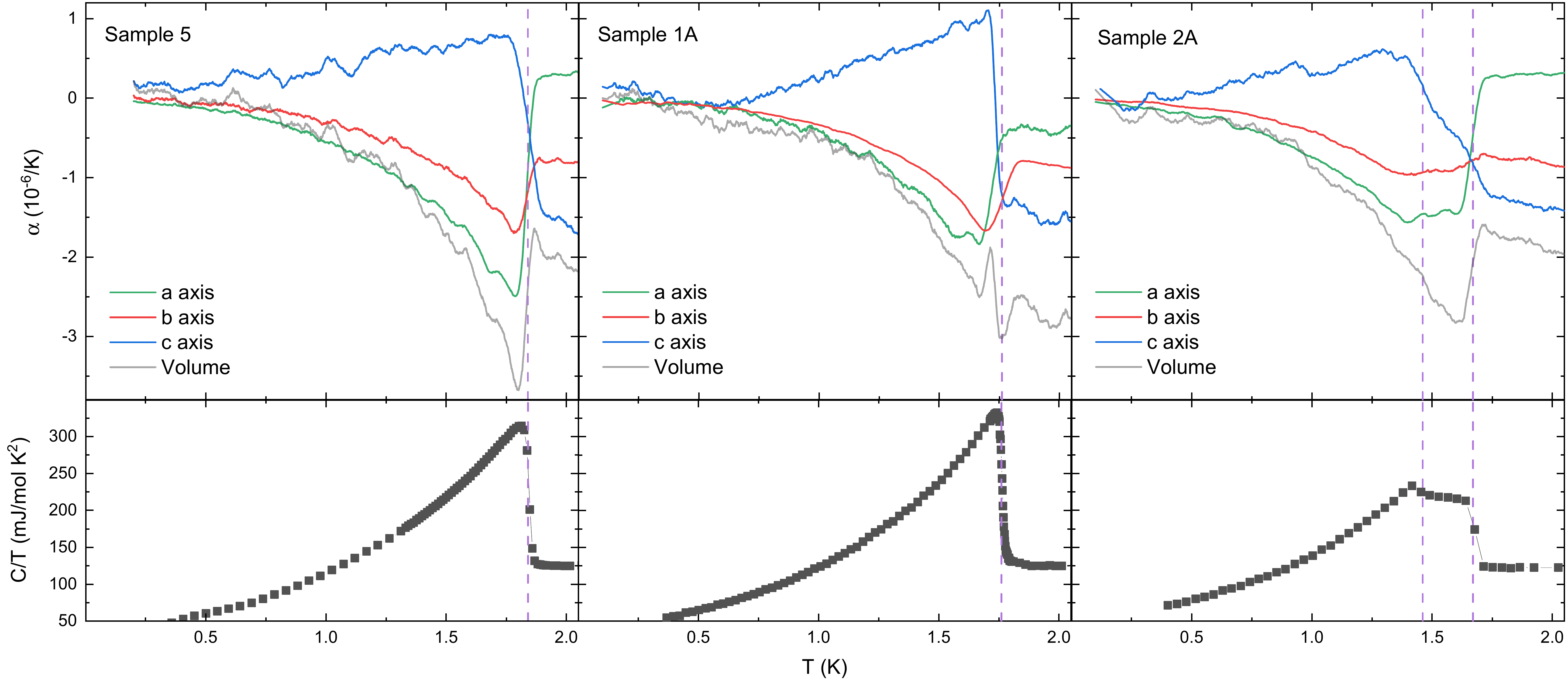}
    \caption{\label{fig:fig1}
        Low temperature thermal expansion (top row) and heat capacity (bottom row) of samples from different growths of UTe\(_2\).
        The purple dashed lines show the transition temperatures determined from heat capacity measurements and their relation to the thermal expansion data.
    }
\end{figure*}

Here, we report thermal expansion, magnetostriction, and heat capacity measurements on a number of UTe$_2$ samples obtained from separate growths to show that growth conditions may lead to two discrete transitions arising from an unusual form of sample inhomogeneity.
In this case, we find clear evidence for two nearby transitions in heat capacity measurements, which are accompanied by jumps in the thermal expansion coefficient.
The relative size of these jumps varies as the measurement is performed on different volumes of the sample, either via thermal expansion or calorimetry measurements.
Importantly, samples with higher T$_c$ only show a single transition at ambient pressure, but all samples measured under applied pressure show at least two detectable superconducting transitions above a threshold pressure.
In samples that show multiple transitions at ambient pressure, a reevaluation of the initial pressure work found that the two transitions at ambient pressure have the same pressure dependence (see erratum to Ref.~\cite{Thomas2020}).

Though we cannot unambiguously rule out the possibility of a multi-component order parameter at ambient pressure for samples with higher T$_c$, our results show no evidence for a second thermodynamic transition below 0.3~GPa.
Because all irreducible representations in UTe$_2$ are one-dimensional, there are only two possibilities for such a scenario: (1)~both the transition temperatures and hydrostatic pressure dependence of the two transitions are accidentally degenerate, or (2)~the lower temperature transition has zero entropy up to 0.3~GPa at which point it can be observed in thermodynamic measurements. Such unlikely scenarios require exceptional fine tuning, which leaves us with the possibility of a nearly vertical line in the pressure-temperature phase diagram.

Samples of UTe\(_2\) were grown using the vapor transport method~\cite{Ran2019,Cairns2020}.
About twenty batches were grown, and representative samples from many different batches were used in this study.
Samples grown at higher temperatures (\textit{i.e.}, 1060\(^\circ{}\)C-1000\(^\circ{}\)C gradient, sample 2) were more likely to show a split transition than samples grown at lower temperatures (\textit{e.g.}, 950\(^\circ{}\)C-860\(^\circ{}\)C gradient, sample 1).
Heat capacity measurements were performed down to \(^3\)He temperatures using the quasi-adiabatic relaxation technique.
Thermal expansion and magnetostriction measurements were performed using a capacitance dilatometer described in Ref.~\cite{Schmiedeshoff2006a} in both \(^4\)He and adiabatic demagnetization cryostats.
All thermal expansion measurements were performed using a slow continuous temperature ramp, whereas all magnetostriction measurements were performed by stabilizing the field to avoid the influence of eddy currents.
Thermal expansion data were corrected by performing a background subtraction of the cell effect under identical thermal conditions.
Ac calorimetry measurements~\cite{Sullivan1968} were performed in a piston clamp pressure cell.
Samples with the same number (1A/1B and 2A/2B) came from the same batch and showed similar zero-pressure heat capacity data.

Figure~\ref{fig:fig1} shows a comparison of thermal expansion and heat capacity between three samples grown under different conditions (for additional samples see Supplemental Fig.~S1).
Sample~5 and sample~1A show a single transition at $T_{c}=1.84$~K and $T_{c}=1.76$~K, respectively.
Sample~2A shows two transitions at $T_{c2}=1.67$~K and $T_{c1}=1.46$~K. 
The difference between these samples highlights the key role of growth conditions on the ambient pressure properties of UTe\(_{2}\).

Importantly, even samples with similar T$_c$ may have different properties.
For instance, samples~5 and 1A have similar heat capacity behavior; however, sample~1A has an unusual negative thermal expansion along the $a$ axis above T$_c$.
Of all the samples measured, sample~1A is the only sample that has $\alpha_a<0$ for $T>T_c$.
As will be discussed below, this may indicate a reduced effect of $a$-axis magnetic fluctuations in this sample.

Volume thermal expansion can be used to determine the pressure dependence of a second order phase transition through the Ehrenfest relation:

\begin{equation}
    \frac{dT_c}{dp}=\frac{\Delta{}\beta{} V_m}{\Delta{}C_p/T_c}.
\end{equation}

\noindent{}Here, \(p\) is pressure, \(\Delta{}\beta{}\) is the jump in volume thermal expansion at the phase transition, and \(\Delta{}C_p\) is the jump in heat capacity.
Because \(\Delta{}C_p\) is always positive, the sign of the pressure dependence is determined by the sign of the jump in volume thermal expansion.
Due to slight temperature offsets when measuring thermal expansion along different axes,  volume thermal expansion jumps were calculated by summing the linear thermal expansion jumps at each phase transition rather than from the volume data.
The results are tabulated in Supplemental Table S1.
Using this relation, sample~5 is expected to have a pressure dependence of approximately $\frac{dT_c}{dp}$=$-0.49(04)$~K/GPa.
This suppression rate agrees well with the pressure dependence of T$_c$ determined from pressure-dependent ac calorimetry measurements (approximately $-0.5$~K/GPa for P~$<$~0.3~GPa).
In contrast, the Ehrenfest relation underestimates the pressure dependence of $T_c$ due to the unusual $a$-axis behavior of sample~1A ($-0.10(04)$~K/GPa).

\begin{figure}[b]
    \includegraphics[width=1\columnwidth]{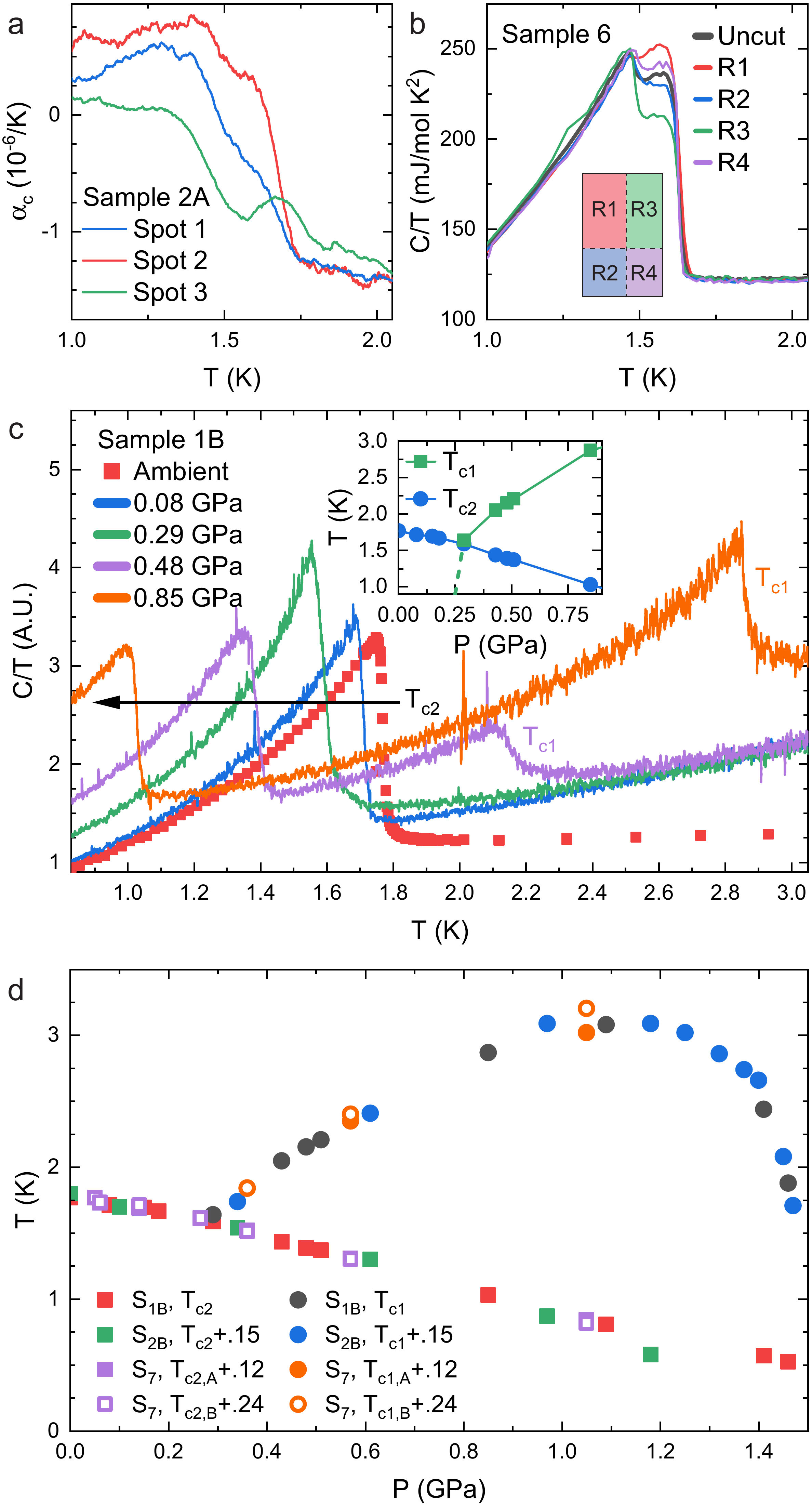}
    \caption{\label{fig:fig2}
        (a)~$c$-axis thermal expansion of sample~2A measured at different locations on the sample.
        (b)~Heat capacity of sample~6 before and after cutting into four quadrants.
        (c)~ac calorimetry measurements of sample~1B.
        Inset shows low-pressure phase diagram.
        (d)~Pressure-temperature phase diagram of all measured samples with T$_c$'s adjusted to match at ambient pressure.
        Samples~2B and 7 were first reported in Ref.~\cite{Thomas2020}.
        Sample 7 has two transitions at ambient pressure (A and B), both of which are tracked as a function of pressure.
    }
\end{figure}

Now we turn to the double transition in sample~2A.
Using the data from Fig.~1, Ehrenfest predicts opposite pressure dependence for the two transitions.
The lower transition has \(\frac{dT_{c1}}{dp}\)=$+0.70(07)$~K/GPa and the higher transition has \(\frac{dT_{c2}}{dp}\)=$-0.65(27)$~K/GPa.
This pressure dependence is most likely incorrect. 
It was recently shown that for samples with two transitions at ambient pressure, the transitions actually have identical pressure dependence (see erratum to Ref.~\cite{Thomas2020}).
The reason for this inconsistency is that the quasi-adiabatic heat capacity measurement probes the entire volume of the sample, but the thermal expansion fixture used here will only probe a local volume of the sample when the sample is measured along its thinnest axis.
For sample~2A, the $c$ axis has a thickness of 300--360~$\mu{}$m compared to 2635~$\mu{}$m and 680~$\mu{}$m for the $a$ and $b$ axes, respectively.

To further unveil this issue, Fig.~2(a) shows the $c$-axis thermal expansion measured on multiple spots of sample~2A.
Spot~1 is the same location that was measured in Fig.~1.
Compared to spot~1, spot~2 has a larger contribution from the higher-temperature transition and a much smaller contribution from the lower temperature transition.
Spot~3 has the opposite weighting between the two transitions.
As a result, the pressure dependence determined from Ehrenfest completely changes based on which location on the sample is used to perform the calculation.
Spot~2 is the only location that predicts a negative pressure dependence for both transitions, in agreement with pressure-dependent ac calorimetry data~\cite{Thomas2020}.

The inhomogeneity of the double transition feature is further demonstrated by the heat capacity measurements shown in Fig.~2(b).
Here, a sample showing two transitions was cut into four quadrants.
The heat capacity of each quadrant was then measured individually.
Remarkably, at temperatures outside the transition region, all four pieces have the same heat capacity.
Near the transition, however, there is a clear difference in the weighting between the two transitions.
Of the four regions, R3 has the largest percentage of the volume containing the lower temperature transition.

\begin{figure}[b]
    \includegraphics[width=1\columnwidth]{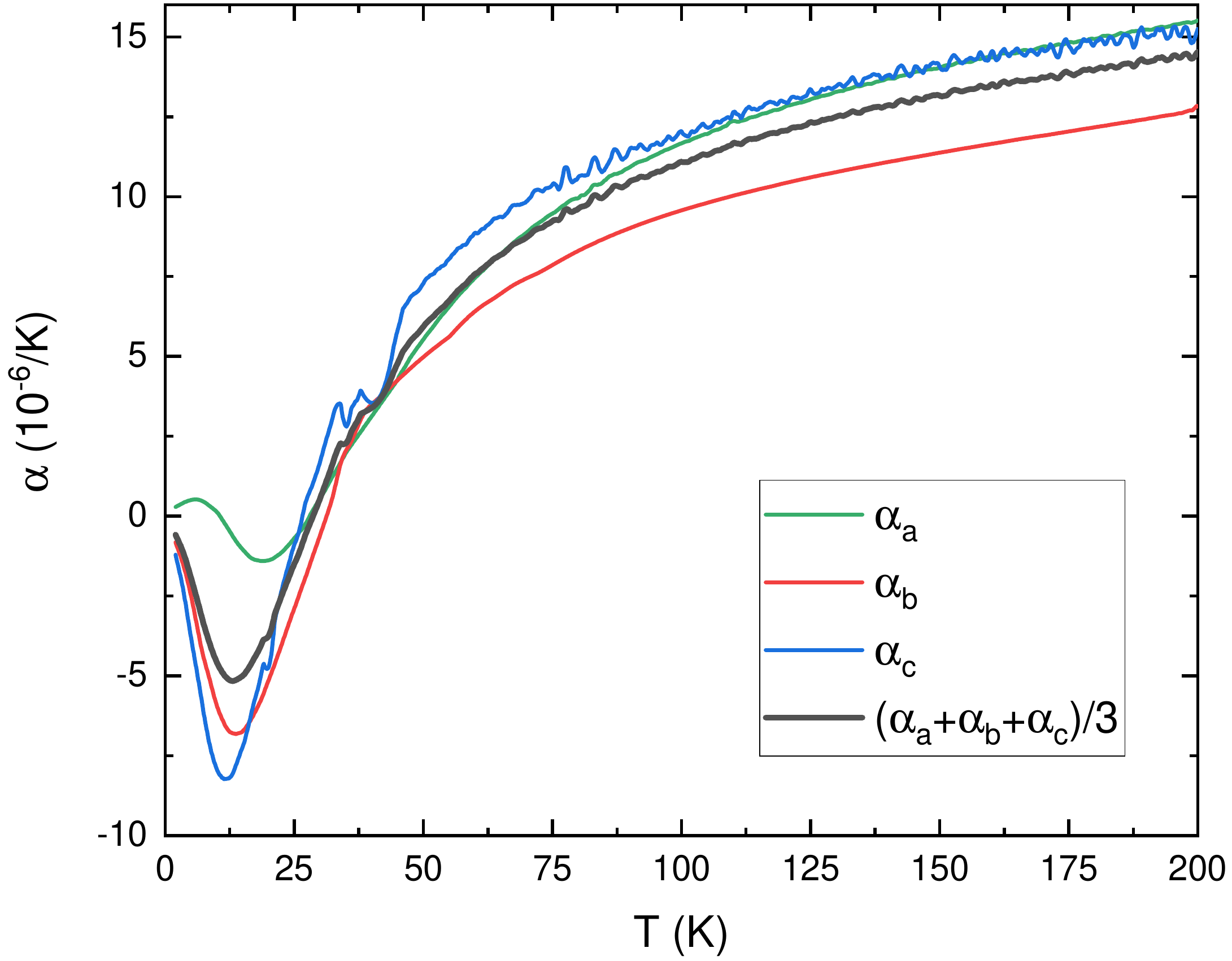}
    \caption{\label{fig:fig3}
        Linear thermal expansion coefficients of sample~2A from 2~K up to 200~K.
        The gray curve shows the volume thermal expansion coefficient divided by three to better fit the scale of linear thermal expansion.
        The features observed near 40~K are due to gas absorption in the insulating washers that are part of the dilatometer cell.
    }
\end{figure}

The reason for the presence of exactly two transitions remains unknown, but our results indicate that the double transition feature at ambient pressure stems from sample inhomogeneity.
Under pressure, however, the two transitions that appear for pressures above 0.3~GPa are an intrinsic feature of UTe$_{2}$
observed in all samples measured by multiple groups~\cite{Braithwaite,Ran2019b,Knebel2020,Aoki2020,Thomas2020}. To confirm this, we performed pressure-dependent ac calorimetry measurements on a sample showing only a single transition at ambient pressure (sample~1B).
The individual heat capacity curves from these measurements are shown in Fig.~2(c), and the pressure-temperature phase diagram is summarized in the inset.
Similar to all other samples measured under pressure~, sample~1B shows clear evidence for two superconducting transitions as pressure is increased beyond 0.3~GPa.
We note that a pressure-temperature phase diagram in which three second order phase transition lines meet at a single point is not thermodynamically allowed except in very unique circumstances~\cite{Yip1991, Braithwaite}, and the dashed line in the inset of Fig.~2(c) is meant to represent this missing transition.
Such a tetracritical point has been extensively investigated in UPt$_3$~\cite{Sauls1994}. 

Remarkably, although samples may have different T$_c$ at ambient pressure, all samples follow the exact same pressure-temperature phase diagram, as shown in Fig.~2(d). This unified diagram is obtained by simply shifting T$_c$ vertically to match a common value at zero pressure (\textit{i.e.}, $T_{c}=$~1.8~K).
This suggests that the main effect of disorder is to suppress T$_c$ and cause a split transition in some samples.
This also reinforces that the splitting of the transition at 0.3~GPa is an intrinsic feature as it is observed in all samples measured to-date.

Thermal expansion to higher temperatures can provide information about the relevant energy scales in the system.
Figure~\ref{fig:fig3} shows the linear thermal expansion for sample~2A measured up to 200~K.
At high temperatures, the thermal expansion is typically dominated by phonons.
Because the non-magnetic analogue ThTe\(_2\) has been reported to have a different crystal structure from UTe\(_2\)~\cite{Koscielski2012}, it is not possible to subtract an independently determined phonon background.
Nonetheless, all three thermal expansion contributions become negative below 30~K indicating a regime wherein the phonon contribution is no longer relevant.
Negative thermal expansion is typically attributed to the Kondo effect, and this temperature is consistent with the Kondo temperature (20--26~K) extracted from scanning tunneling spectroscopy measurements~\cite{Jiao2019}.
Expansion along the a-axis shows a third energy scale, switching again from negative to positive at 11~K.
This is likely due to the presence of strong magnetic fluctuations along the $a$~axis, in agreement with previous reports~\cite{Ran2019,Sundar2019,Tokunaga2019}.
While samples 2--5 all exhibit positive thermal expansion along the $a$~axis just above the highest temperature superconducting transition, sample~1A has negative thermal expansion along $a$.
This may point to a difference in the strength or type of magnetic fluctuations along the $a$~axis that is also influenced by differences in growth conditions. 

\begin{figure}[t]
    \includegraphics[width=.95\columnwidth]{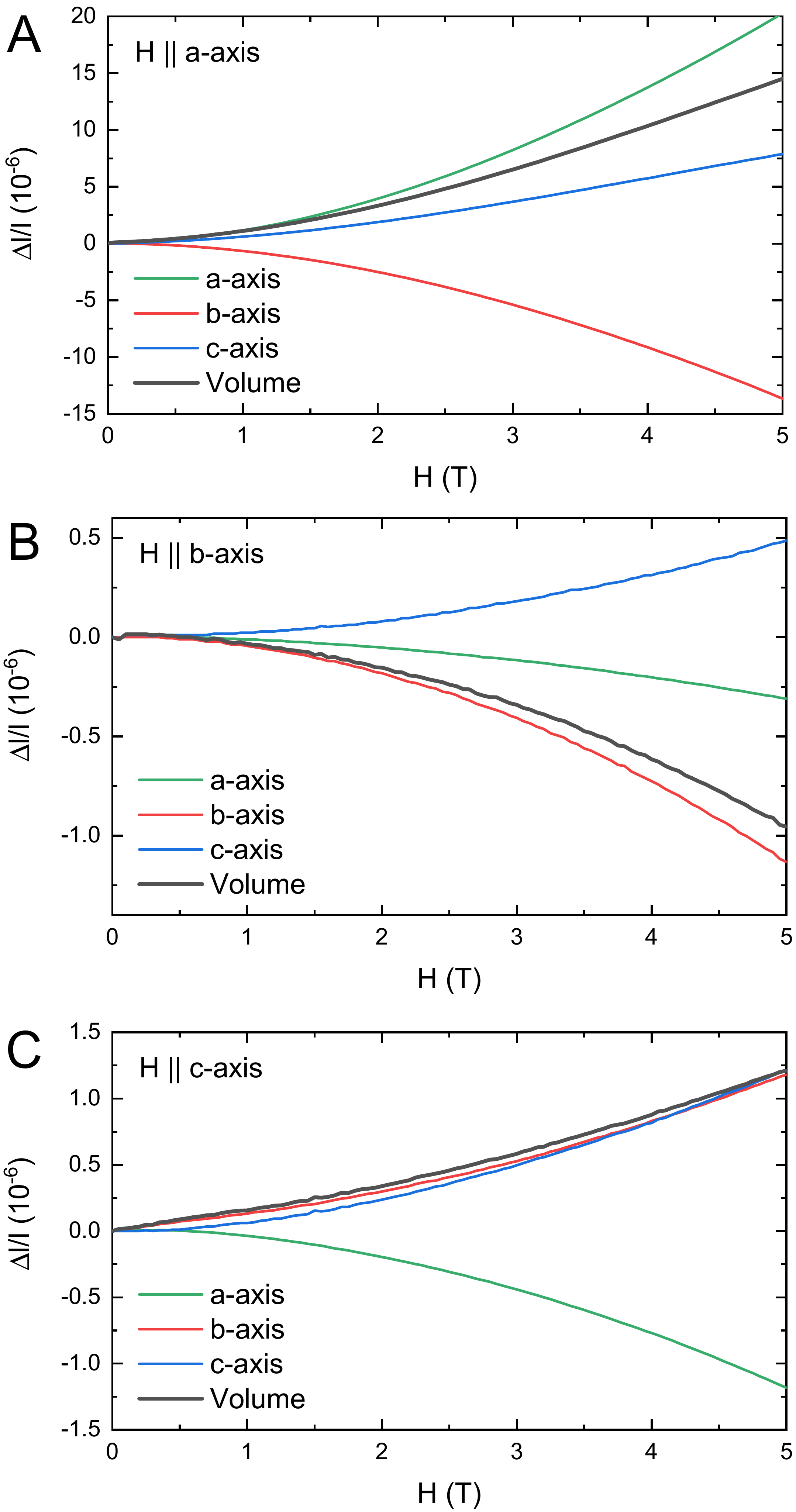}
    \caption{\label{fig:fig4}
        Linear and volume magnetostriction for magnetic fields applied along the three principal axes of sample~3. All data were obtained at 2.0~K.
    }
\end{figure}

Figure~\ref{fig:fig4} shows the longitudinal and transverse magnetostriction measured at 2~K on sample~3 along the principal crystallographic directions.
Note that the response for fields parallel to the $a$~axis is an order of magnitude larger than along the other axes.
This means that even a small field component parallel to the $a$~axis will significantly affect measurements when applying field along other directions.
For longitudinal measurements, the sample was aligned to less than one degree using Laue diffraction.
For transverse measurements, the rotation of the sample in the dilatometer cell was performed manually so the alignment errors may be up to five degrees and introduce an error in measurements for fields perpendicular to the a-axis.

Volume magnetostriction can be used to determine the pressure dependence of the magnetic susceptibility via Maxwell's relation~\cite{Fawcett1970}:

\begin{equation}
    {\left(\frac{\partial{\chi{}}}{\partial{P}}\right)}_{H,T}\propto{}-{\left(\frac{\partial{}V}{\partial{}H}\right)}_{P,T}
\end{equation}

At ambient pressure, the $a$~axis is the easy magnetic axis and the $b$~axis is the hard magnetic axis~\cite{Ran2019}.
Importantly, the volume magnetostriction for fields parallel to the $a$~axis indicates a relatively large negative pressure dependence of the $a$~axis susceptibility.
This is consistent with a recent tight-binding model for UTe\(_2\), which found a large initial decrease in susceptibility along the $a$~axis coupled with a change in the fluctuations from ferromagnetic to antiferromagnetic~\cite{Ishizuka2020}.
Further, the volume increase for fields along the $c$~axis taken with the decrease for fields along the $b$~axis suggests the possibility that the hard magnetic axis changes from the $b$~axis to $c$~axis.
This has previously been claimed based on the fact that H\(_{c2}\) becomes largest along the $c$~axis near 1.5~GPa~\cite{Knebel2020}.
More recently, it was experimentally confirmed via susceptibility measurements under pressure~\cite{Li2021}.
The magnetic interactions at high pressure are quite different from those at low pressure, which explains the emergence of two magnetic transitions and antiferromagnetic order.
In fact, the $b$ axis becomes the easy axis in the magnetically ordered state~\cite{Li2021}.
We also highlight the possibility that samples from different batches may exhibit different magnetic properties even at ambient pressure.
This follows from the fact that sample~1A has a different sign of $\alpha_a$ just above T$_{c2}$ compared to samples 2--4, as noted above.
Thus, it is critical to fully characterize each single crystal of UTe$_2$. 

In conclusion, the combination of thermal expansion and heat capacity shows evidence for two superconducting transitions at ambient pressure only in some UTe$_{2}$ samples.
Our results indicate that the double transition is due to different T$_c$'s in spatially separated volumes of the sample that are inhomogenously distributed. This in turn implies that these two transitions do not arise from a multi-component order parameter.
If UTe$_2$ possesses a multi-component order parameter at ambient pressure, it must be detected through other means, as evidence for two transitions in thermodynamic data is misleading in this material.
Nonetheless, all samples measured to-date show clear evidence for a splitting of T$_c$ under pressure, which strongly 
suggests that this feature is intrinsic.
Our magnetostriction data also agree with recent theoretical and experimental work that argues for a change in the nature of the magnetic interactions under pressure.
Our results reveal that subtleties in sample growth play a large role in both superconductivity and magnetic fluctuations in UTe$_{2}$. Detecting the lower temperature transition for pressures below 0.3~GPa will play a major role in illuminating the nature of the superconducting state at ambient pressure.
The origin of the sample dependence in UTe$_{2}$ may be related to structural changes, strain, or stoichiometry variations, and this topic also needs to be further investigated in the near future.

\begin{acknowledgments}
We would like to thank N. Harrison, M. Jaime, and R. M. Fernandes for useful discussions as well as L. Gonzales for assistance running experiments.
Thermal expansion and magnetostriction measurements were supported by the U.S. Department of Energy, Office of Basic Energy Sciences, Division of Materials Science and Engineering project ``Quantum Fluctuations in Narrow-Band Systems.''
Sample synthesis at Los Alamos was performed with support from the U.S. Department of Energy, Office of Science,  National Quantum Information Science Research Centers, Quantum Science  Center.
Pressure-dependent measurements were supported by the Laboratory Directed Research and Development program 20210064DR.
C. Stevens and A. Huxley acknowledge support from UK-EPSRC grant EP/P013686/1.
F.B. Santos was supported by FAPESP under grants No. 2016/11565-7 and 2018/20546-1.

\end{acknowledgments}

\bibliography{lib.bib}

\end{document}


\title{Supplemental Information for ``Spatially inhomogeneous superconductivity in UTe$_{2}$''}

\author{S. M. Thomas$^{1}$, C. Stevens$^{2}$, F. B. Santos$^{3}$,  S. S. Fender$^{1}$, E. D. Bauer$^{1}$, F. Ronning$^{1}$, J. D. Thompson$^{1}$, A. Huxley$^{2}$, and P. F. S. Rosa$^{1}$}
\affiliation{
$^{1}$  Los Alamos National Laboratory, Los Alamos, New Mexico 87545, U.S.A.\\
$^{2}$ School of Physics and Astronomy, University of Edinburgh, Edinburgh, UK.\\
$^{3}$ Escola de Engenharia de Lorena, Universidade de Sao Paulo (EEL-USP), Materials Engineering Department (Demar), Lorena, Sao Paulo, Brazil.}
\date{\today}

\maketitle

\subsection{Additional thermal expansion and heat capacity data}
Even though sample~3 only shows a single transition, it is clear that sample~3 is tuned to slightly higher pressures compared to sample~1A.
The heat capacity anomaly is smaller, which implies a different balance between the superconducting transitions, and the lower $T_{c}$ is consistent with the magnitude obtained at pressures nearby the crossing point found in sample~2B ($\sim{}0.2$~GPa).
In addition, the largest thermal expansion jump in sample~3 occurs along $a$, whereas in sample~1A the $c$~axis thermal expansion jump is the largest.
In sample~4, T$_{c1}$ appears to occur at a higher temperature than T$_{c2}$, which suggests that this sample is tuned beyond the crossing point in the pressure-temperature phase diagram.
Although sample~4 shows two transitions in heat capacity, it is not possible to resolve the higher temperature transition in thermal expansion.
It is likely that the thermal expansion anomaly is below the noise level of the measurement due to the small dimensions of the sample.
\renewcommand{\thefigure}{S1}
\begin{figure*}[h]
    \includegraphics[width=0.9\textwidth]{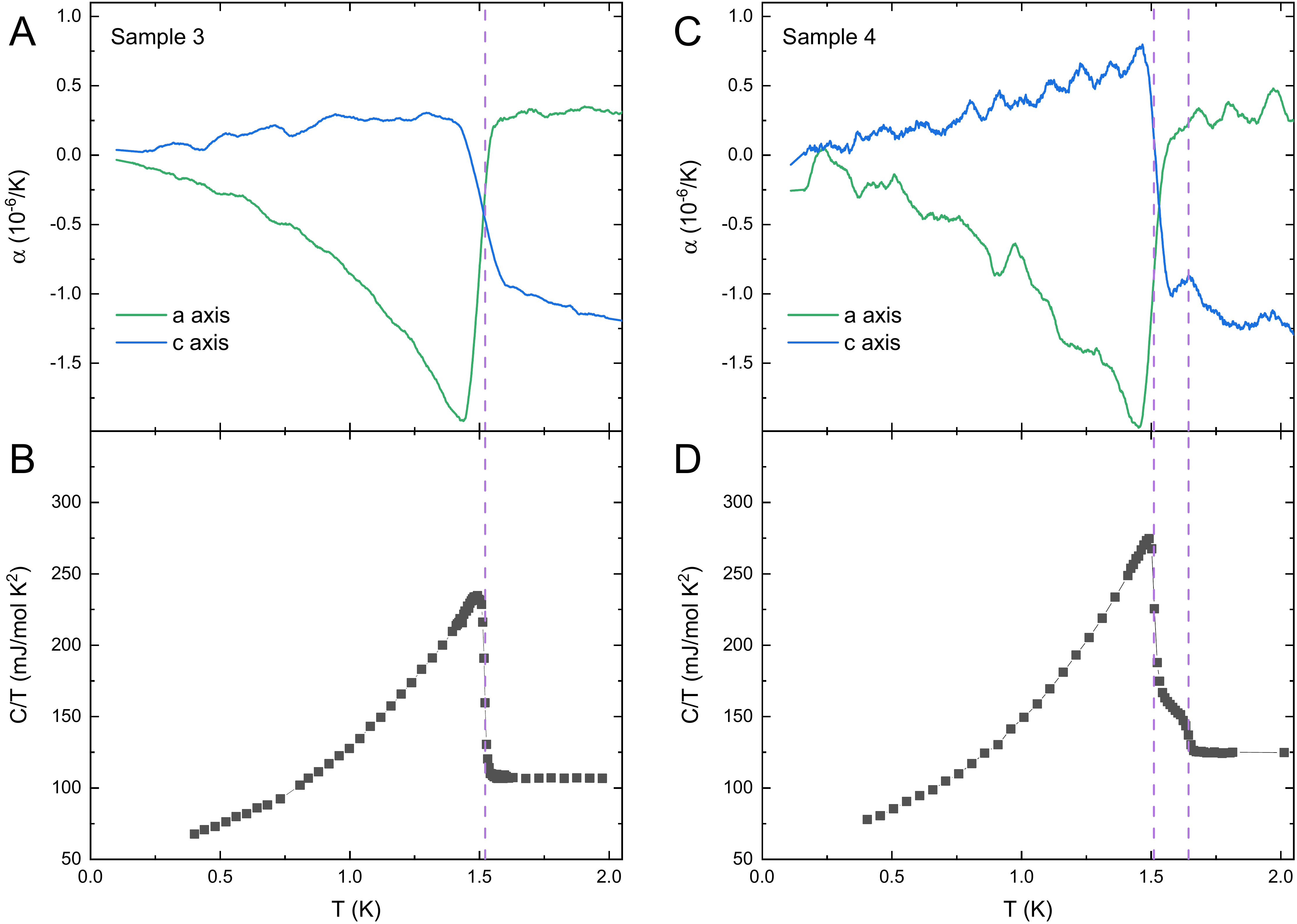}
    \caption{\label{fig:sfig1}
        (A)--(D) Low temperature thermal expansion (top row) and heat capacity (bottom row) of samples from different growths of UTe\(_2\).
        The purple dashed lines show the transition temperatures determined from heat capacity measurements extended up to the thermal expansion data.
        By comparison with all other samples, the lower temperature transition appears to be T$_{c2}$.
    }
\end{figure*}

\newpage
\renewcommand{\thetable}{S1}
\subsection{Tabulated values}
Thermal expansion and heat capacity jumps were calculated by performing a linear fit just above and below the transition and calculating the difference between the fits at the midpoint of the transition. Error bars in thermal expansion jumps were estimated by considering two main sources of error coming from uncertainty in the length of the sample and measurement noise in alpha.
\begin{table*}[h!]
    \centering
    \begin{tabular}{| r | c | c | c | c | c | c | c |}
        \hline 
        & T (K) & $\Delta\alpha_a$ ($10^{-6}$/K) & $\Delta\alpha_b$ ($10^{-6}$/K) & $\Delta\alpha_c$ ($10^{-6}$/K) & $\Delta\beta$ ($10^{-6}$/K) & $\Delta{}C/T$ (J/mol K$^2$) & $\frac{dT_c}{dP}$ (K/GPa) \\
        \hline
        \multicolumn{1}{|l|}{Sample 1A} & & & & & & &  \\
        $T_{c}$ & $1.76$ & $-1.56(14)$ & $-1.22(03)$ & $+2.39(06)$ & $-0.39(15)$ & $0.219$ & $-0.10(04)$ \\
        \hline
        \multicolumn{1}{|l|}{Sample 2A} & & & & & & & \\
        Spot 1 & & & & & & & \\
        $T_{c1}$ & $1.46$ & $-0.24(05)$ & $-0.21(06)$ & $+0.74(09)$ & $+0.29(12)$ & $0.024$ & $+0.65(27)$ \\
        $T_{c2}$ & $1.67$ & $-1.64(05)$ & $-0.18(06)$ & $+0.51(09)$ & $-1.31(12)$ & $0.099$ & $-0.70(07)$ \\
        Spot 2 & & & & & & & \\
        $T_{c1}$ & & & & $+0.25(09)$ & $-0.20(12)$ & & $-0.45(27)$ \\
        $T_{c2}$ & & & & $+1.19(09)$ & $-0.63(12)$ & & $-0.34(07)$ \\
        Spot 3 & & & & & & & \\
        $T_{c1}$ & & & & $+1.03(09)$ & $+0.58(12)$ & & $+1.22(27)$ \\
        $T_{c2}$ & & & & $+0.25(09)$ & $-1.57(12)$ & & $-0.85(07)$ \\
        \hline
        \multicolumn{1}{|l|}{Sample 3} & & & & & & & \\
        $T_{c}$ & $1.51$ & $-2.40(05)$ & N/M & $+1.15(08)$ & N/A & $0.123$ & N/A \\
        \hline
        \multicolumn{1}{|l|}{Sample 4} & & & & & & & \\
        $T_{c1}$ & $1.64$ & N/D & N/M & N/D & N/A & $0.021$ & N/A \\
        $T_{c2}$ & $1.46$ & $-2.22(23)$ & N/M & $+1.68(18)$ & N/A & $0.118$ & N/A \\
        \hline
        \multicolumn{1}{|l|}{Sample 5} & & & & & & & \\
        $T_{c}$ & $1.84$ & $-2.92(05)$ & $-1.09(07)$ & $+2.28(11)$ & $-1.73(14)$ & $0.189$ & $-0.49(04)$ \\
        \hline
    \end{tabular}
    \caption{List of parameters. N/M -- not measured, N/D -- not detected, and N/A -- not applicable.
    The quantity in parenthesis indicates the estimated uncertainty in the last two digits.
    }
    \label{table:1}
\end{table*}

The length of the sample along the direction of the thermal expansion measurement is listed below.
The main effect of shorter length is to increase the noise in alpha.
$\Delta{}L\propto{}\Delta{}C$ and the measurement noise of $C$ is mostly independent of sample length ($L$ is sample length and $C$ is the measured capacitance value of the dilatometer).
As $\alpha=\frac{1}{L}\frac{\mathrm{d}L}{\mathrm{d}T}$, dividing by a larger L reduces the noise of the measurement.

\renewcommand{\thetable}{S2}
\begin{table*}[h!]
    \centering
    \begin{tabular}{| r | c | c | c |}
        \hline 
        & a ($\mu$m) & b ($\mu$m) & c ($\mu$m) \\
        \hline
        Sample 1A & 310 & 2106 & 1313 \\
        \hline
        Sample 2A & 2635 & 680 & \\
        Spot 1 & & & 360 \\
        Spot 2 & & & 300 \\
        Spot 3 & & & 300 \\
        \hline
        Sample 3 & 2405 & & 900 \\
        \hline
        Sample 4 & 930 & & 330 \\
        \hline
        Sample 5 & 2667 & 1707 & 330 \\
        \hline
    \end{tabular}
    \caption{Measured sample length along the direction of thermal expansion measurement.
    }
    \label{table:2}
\end{table*}

\newpage

\subsection{Thermal expansion measurement details}

The dilatometer makes contact with the sample on a flat platform on one edge and a small, rounded tip on the opposite edge.
Because of this, the thermal expansion measurement may only probe a volume of the sample underneath the rounded tip, especially when the sample is thin.
This is shown in Fig.~S2 below for two different sample geometries.
The sample is pressed into the contacting trip, which causes a deflection of the CuBe spring.
The outside of the spring is fixed in place.
This induces a small stress in the sample that indicates the region of the sample that is influencing the position of the contacting tip.
The deflection of the contacting tip is what determines the measured length change of the sample.
For a long sample, as shown in the middle image, there is nearly uniform stress in the sample.
For a thin sample, only the area directly under the contacting tip experiences any stress (bright red regions are unstressed).
The result is that measurements along smaller lengths will be more dependent on where the sample is contacted.

\renewcommand{\thefigure}{S2}
\begin{figure*}[h]
    \includegraphics[width=1\textwidth]{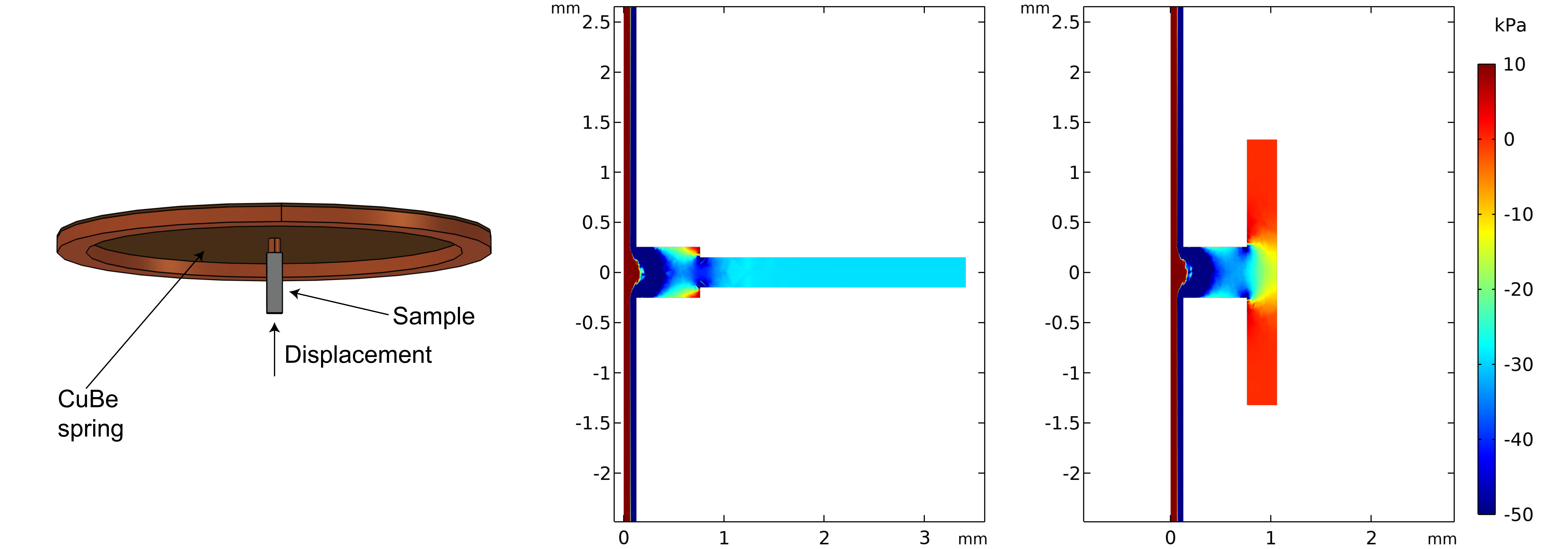}
    \caption{\label{fig:sfig2}
        Stress induced in sample by copper beryllium spring due to thermal expansion of the sample.
    }
\end{figure*}